# Low field magnetocaloric effect and heat capacity of A-site ordered NdBaMn$_2$O$_6$ manganite

A.M. Aliev*[1], A.G. Gamzatov**[1], V.S. Kalitka[2], A.R. Kaul[2]

[1]Amirkhanov Institute of Physics, Daghestan Scientific Center of RAS, 367003, Makhachkala, Russia

[2]Moscow State University, 119899, Moscow, Russia

E-mail: *lowtemp@mail.ru; **gamzatov_adler@mail.ru

**Annotation**

The magnetocaloric effect (MCE) in the A-site ordered manganite NdBaMn$_2$O$_6$ is studied. The MCE in this compound has an anomalous behavior. In low magnetic fields, the abrupt transitions between direct and inverse magnetocaloric effect are observed. In a relatively strong magnetic field $H$=11 kOe the direct and inverse effects are observed only at cooling, while the heating mode reveals only an inverse MCE. The value of the MCE (-$\Delta S$=0.7 J/kg K and $\Delta S$= 1.02 J/kg K for $\Delta H$=11 kOe) does not reach high values, but the proximity of the effects occurring at room temperatures expects the use of both effects in the magnetic cooling technology.

**Introduction**

A-site ordered $R$BaMn$_2$O$_6$ manganites ($R$-rare earth) have attracted the attention of researchers because of their unique physical properties. The most significant structural feature of these materials is that the MnO$_2$ planes are located between $RO$ and BaO layers, which have significantly different sizes, resulting in highly distorted MnO$_6$ octahedra. This implies that the structural and physical properties of $R$BaMn$_2$O$_6$ can not be explained using only the tolerance factor, as for the cation-disordered $R_{0.5}$Ba$_{0.5}$MnO$_3$ manganites [1]. Furthermore the distortion may have an additional effect on the competition of the charge, orbital, spin and lattice degrees of freedom. Phase transition temperatures of A-site ordered manganites strongly depend on the rare-earth element (see phase diagrams in [1-3]) and on conditions of the sample preparation [4].

In recent years the researchers have directed their attention to study of materials with giant magnetocaloric effect. This is conditioned by the possibility of practical application of such materials in solid-state magnetic refrigerators. The maximum possible values of MCE in simple magnetic materials are not sufficient to create an effective magnetic refrigerator [5]. Therefore, research efforts are pressing for search and study of magnetic materials, where other types of ordering (structural, charge, orbital) are observed. These orderings may contribute additional effect to the overall MCE. In this sense, the A-site ordered manganite NdBaMn$_2$O$_6$ is of great interest. In this compound the ferromagnetic, antiferromagnetic transitions and orbital ordering practically merge into a single point [1-3, 6]. Because of proximity of the magnetic phase transitions accompanied by structural and orbital ordering, it can be expected highly complex behavior of the MCE in this compound in the dependence on an applied magnetic field and a direction of change in the sample temperature.

This work presents the results of the MCE explored in different magnetic fields using the direct and indirect techniques. Direct measurements of the MCE were performed by a modulation method [7], and indirectly the MCE was studied by means of heat capacity measurements. The heat capacity was measured by a.c. -calorimetry, magnetization was gauged on a vibrating magnetometer, and a four-probe method was used for estimation of the resistivity.

**Results and Discussion**

NdBaMn$_2$O$_6$ ceramic samples were prepared by chemical homogenization. Ashfree paper filters saturated with a stoichiometric mixture of nitrate solutions of Nd, Ba and Mn and dried at 100 ° C. Next, they were burned and calcined at 600 ° C. Received powder was compressed into pellets and annealed at 1100 °C in Ar stream for 20 hours. As the result of sintering was obtained the manganite Nd$_{0.5}$Ba$_{0.5}$MnO$_3$ without A-cations ordering. Next, the sample was annealed in a sealed tube at 1100 ° C and $P(O_2) = 10^{-20}$ atm. also during 20 hours. Oxygen partial pressure was gettered by Fe / FeO, located at 800 °C. At the same time there happened the ordering of A-sublattice cations and the formation of $NdBaMn_2O_5$ manganite. The sample was annealed at 500 °C under $O_2$ stream for 5 hours to fill the oxygen vacancies of $NdBaMn_2O_{5.97(1)}$ manganite (index of oxygen was refined by a method of iodometric titration). The results of X-ray diffraction displayed a single phase in the obtained sample with unite cell parameters $a=b=3.8987(3)$, $c=7.7251(9)$ (Group P4mm). From the powder XRD data (see Fig.1) for ordered $NdBaMn_2O_6$ sample we can see, that diffraction peaks of cubic perovskite subcell are spited due to tetragonal distortion. Also near 11° $2\theta$ there could be found a (00½) peak, which proves the superlattice formation. All peaks on this pattern belong to the manganite phase. Cell parameter for oxidized phase $NdBaMn_2O_6$ were refined from powder XRD data in program Jana2006 using lattice constrain full profile refinement.

Figure 2 shows the magnetization in a wide temperature range. Temperature dependence of magnetization does not point strictly to the paramagnetic - ferromagnetic (PM-FM) phase transition. Rather, a sharp increase in magnetization, starting at $T \sim 310$ K, indicates the rise in ferromagnetic correlations [3]. Typically, the FM-PM transition in manganites is accompanied by metal-insulator transition what is not in point of fact seen in the temperature dependence of resistivity (Fig. 3), i.e. a behavior of $\rho(T)$ is semiconductive. The magnetization decreases sharply below 290 K. It may be connected both with sharp change in the sample volume resulting in magnetization decreasing and in consequence of the transition to the antiferromagnetic state. The neutron diffraction experiment shows that below this temperature an A-type antiferromagnetically ordered state appears in the system [3], i.e. a structure, where the *ab* planes are ordered ferromagnetically, and along the *c* axis these planes are ordered antiferromagnetically. A-type antiferromagnetic ordering always is accompanied by some orbital ordering, in this case it is $d_{x2-y2}$ orbital ordering [3]. The character of MCE behavior can be traced by differential $M(T)$ curve with respect to the temperature. The inset in Figure 2 shows that the direct (i.e., $\Delta T > 0$, $\Delta S < 0$) and inverse ($\Delta T < 0$, $\Delta S > 0$) magnetocaloric effects with a maximum at $T = 292$ K and $T = 287$ K, respectively, must be observed in a narrow temperature range for this composition. Considering the proximity of magnetic phase transitions accompanied by structural and orbital ordering, one can expect a complex behavior of the MCE in this composition in low and strong magnetic fields and depending on direction of the sample temperature changes.

The heat capacity was measured at zero-field and in the field of 11 kOe (Fig. 3 and 4) to study the MCE in strong magnetic fields. Heat capacity in the heating mode was measured after cooling the sample in zero magnetic field (ZFC). At the multicritical point a heat capacity anomaly is observed as a peak, narrow, but large in magnitude. Such anomalies are typical for the first order phase transitions. It is confirmed by the thermal hysteresis of the heat capacity; the heat capacity peak takes place at $T=290.1$ K in the heating mode, and in the cooling mode occurs at $T=280.3$ K, with peaks different in magnitudes. The temperature dependence of resistivity (Fig.3) also reveals the weak anomalies at temperatures corresponding to maxima of heat capacity and magnetization, at heating the anomaly is invisible in a scale of the Figure.

Application of a magnetic field leads to a shift of the peak of the specific heat towards the lower temperatures, what is usually characteristic to the antiferromagnetic phase transition. A change in modes of the magnetic fields displays the peaks to different values: at heating – by 1 K, and at cooling – more than by 2 K (Fig.4). These differences can lead to differences in the behavior of the MCE in these modes. MCE was calculated using the heat capacity data by the formula $\Delta S = \int_0^T \left( (C_{P,0} - C_{P,H})/T \right) dT$.

The estimated MCE are shown in Figure 5. In cooling mode there are negative change entropy at high temperatures, with a peak at $T$=286 K, what is typical for ferromagnetics. Further decreasing in temperature sharply changes a sign of the MCE, and when $T$=278.7 K here occurs the maximum of inverse (antiferromagnetic) MCE. But in the heating mode we observe completely other behavior of the MCE, namely, there is only an inverse effect with peak at $T$=289.6 K. Further increase of the temperature does not result to change of the MCE sign.

The value of $\Delta S$ in the studied compound NdBaMn$_2$O$_6$ does not reach high values. The magnitude of the direct effect $\Delta S$=-0.34 J / mol K (or -0.7 J/kg K) at magnetic field change of 11 kOe. It is noteworthy that the obtained values of the direct MCE are in good agreement with the results of Zhang and et al. [8] ($\Delta S$=-1.2 J/kg K for $\Delta H$=20 kOe), in spite of the large difference in the temperatures of phase transitions of the studied samples. Here we note that the Neel point in the sample studied in [8] was detected at $T_N$=210 K, that did not conform to the phase diagrams of the A-site ordered manganites [1-3].

Differences in the behavior of the MCE in heating and cooling modes in NdBaMn$_2$O$_6$ can be understood by taking into account the magnetic properties of this material. Ferromagnetic correlations are observed in the high temperature region in NdBaMn$_2$O$_6$. When applying an external magnetic field, these correlations contribute to the direct MCE, and with cooling the magnitude of the effect increases at that, because decrease in the disordering effect of thermal energy. We observe an inverse magnetocaloric effect when the sample turns to AFM state. At heating, there is a different picture. At low temperatures, AFM is the ground state of NdBaMn$_2$O$_6$ and inverse MCE occurs. But the direct effect is not detected above the Neel point as the magnetic field applied still in the AFM state suppresses the fluctuations failing the appearance of FM correlations. Otherwise it may be explained as freezing of AFM state by the magnetic field. Thus, in heating mode, we observe the typical antiferromagnetic behavior of MCE, i.e. only the inverse effect near the antiferromagnetic transition. The magnetic field will suppress the fluctuations, of course, if it were initially applied in the high temperature phase. But in this phase the ferromagnetic correlations exist prior the application of the field, and the external magnetic field will strengthen the existing correlations preventing formation of new regions, so there is the direct MCE in this case. If such a behavior picture is true for MCE, the direct and inverse effects should be observed both in heating and cooling modes in low fields when the magnetic field cannot freeze the AFM state.

The measurement results of the MCE in weak magnetic fields are presented in Fig. 6. As follows from the figure in this case the direct and inverse MCE are observed both in heating and cooling modes. The heating and cooling modes reveal different indications in low fields. Thermal hysteresis for the direct MCE equals to 7 K and $T$=10 K for the inverse one. The effect magnitude also depends on the direction of temperature change: a value of the inverse effect is greater in the heating cycle, and a value of the direct effect rises in the cooling cycle. The width of transition regions is different at heating and cooling too. In former mode the temperature width between the peaks of the inverse and direct MCE is about 3 K, while in latter mode is about 6 K, i.e. the MCE sign changes more dramatically in the heating mode. MCE temperature peaks slightly depend on a magnetic field, a significant change of $T_{peak}$ observed only for the inverse effect in the cooling mode. In low fields the peaks temperatures for inverse MCE mainly coincide with the maxima temperatures of the specific heat in zero field ($T(C_{p\ peak})$ = 290.1 K, $T(-\Delta T_{\ peak})$ = 289.8 K in the heating mode; $T(C_{p\ peak})$ = 280.3 K, $T(-\Delta T_{\ peak})$ ~ 280 K in the cooling mode).

The low-field values can be extrapolated to the strong fields to compare the MCE values in $NdBaMn_2O_6$ with effects in other magnetocaloric materials. At $\Delta H$=750 Oe the MCE value is equal to $\Delta T$=0.04 K. Linear extrapolation leads to a value $\Delta T$=0.58 K for $\Delta H$=11 kOe. On the other hand, the estimation of $\Delta T$ using the heat capacity data and the relation $\Delta T = \Delta S(T/C_p)$ deduces $\Delta T$=0.9 K ($\Delta H$=11 kOe). Taking into account that the estimation of the MCE was carried out in the first order phase transition region, where the error estimate from heat capacity data can achieve significant values, such a coincidence of the obtained data can be considered satisfactory. The MCE value in $NdBaMn_2O_6$ is not large in comparison with classical magnetocaloric materials. However, the transitions between direct and inverse magnetocaloric effects observed at room temperature are of interest from the standpoint of studying the interaction of different orderings and the effect of external parameters (magnetic field, temperature) on them. Also this may be important from the practical application viewpoint; a refrigeration unit of appropriate design can be used in one cycle both the direct and inverse effects virtually doubling the cooling efficiency of magnetic refrigerators. As well a tune of these effects by changing of applied magnetic field or by different modes of temperature change may have the importance. Abrupt temperature transitions between direct and inverse magnetocaloric effects are observed in many materials including manganites [9]. But in $NdBaMn_2O_6$ the situation is different thereby that the sequence of these transitions depends on a mode of the temperature change and a magnitude of the external magnetic field tension. Application of the stronger magnetic field must induce AFM-FM transition. Thus a sequence of the magnetocaloric effects: inverse and direct - inverse – direct must occur in the heating mode depending on the applied magnetic field (weak - medium - high). Additional researches in middle and strong magnetic fields are required for complete understanding of the MCE nature in $NdBaMn_2O_6$. Similar picture in MCE and heat capacity behavior we have observed for $PrBaMn_2O_6$ [10, 11], with the difference that for $PrBaMn_2O_6$ compound the AFM – FM transitions are strongly extended along the temperature, therefore an interaction influence of AFM – FM orderings on MCE is slacked.

**Acknowledgements**

This work was partly supported by RFBR (Grant Number 09-08-96533, 11-02-01124) and the Physics Department of RAS.

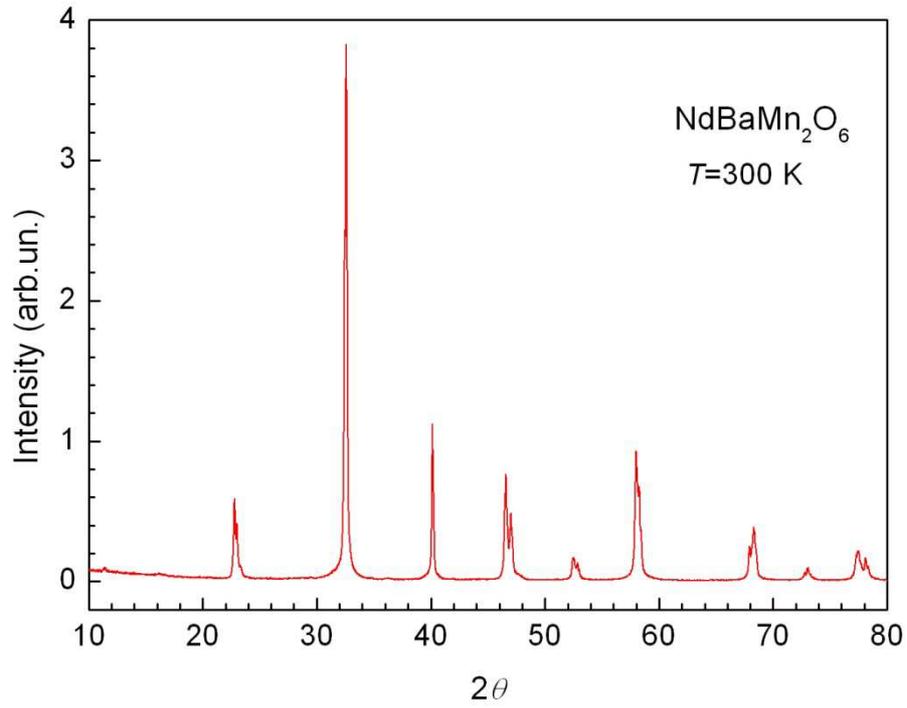

Figure 1. Powder XRD pattern of NdBaMn$_2$O$_6$ manganite.

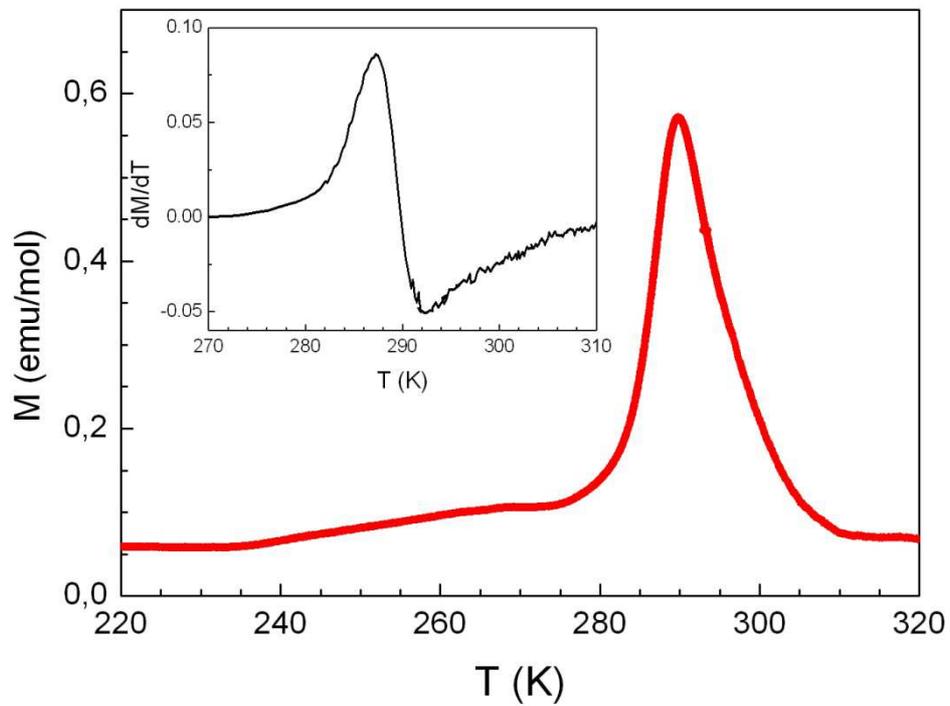

Figure 2. Magnetization of NdBaMn$_2$O$_6$ in heating mode.

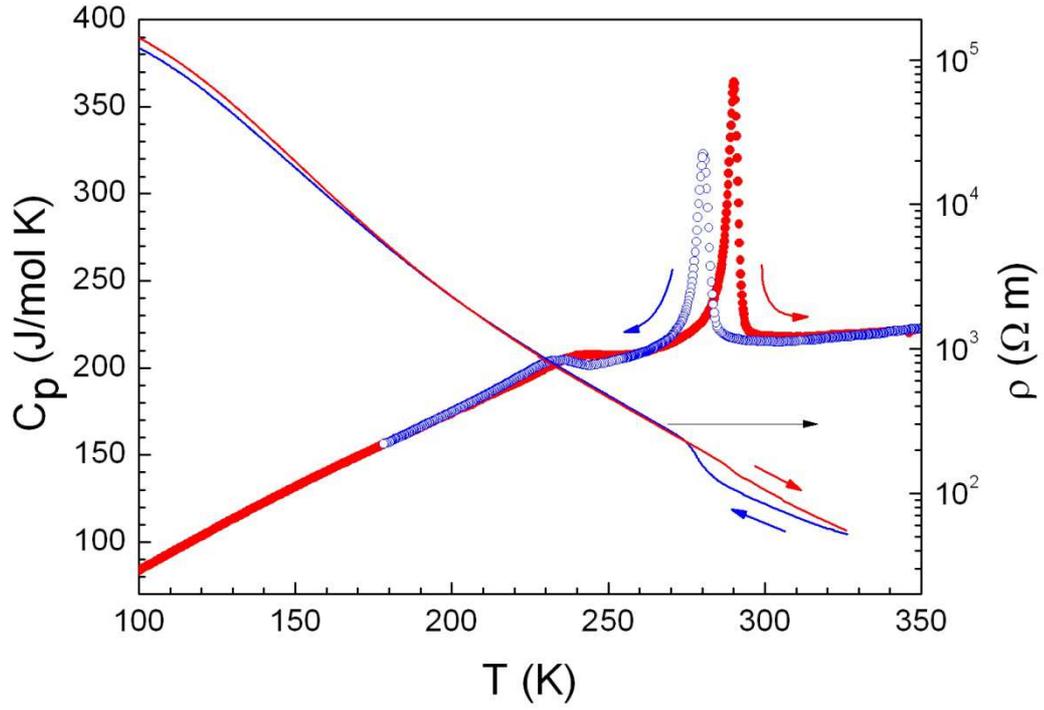

Figure 3. Temperature dependences of NdBaMn$_2$O$_6$ heat capacity and resistivity in heating and cooling modes.

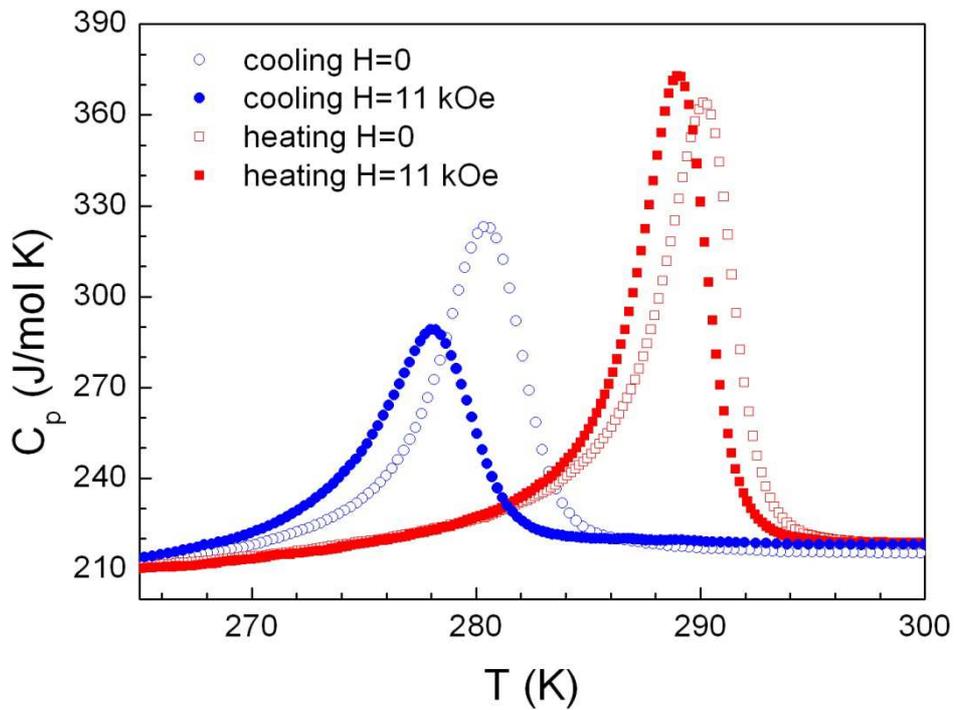

Figure 4. Temperature dependences of the heat capacity in phase transition region at 11 kOe and zero-field.

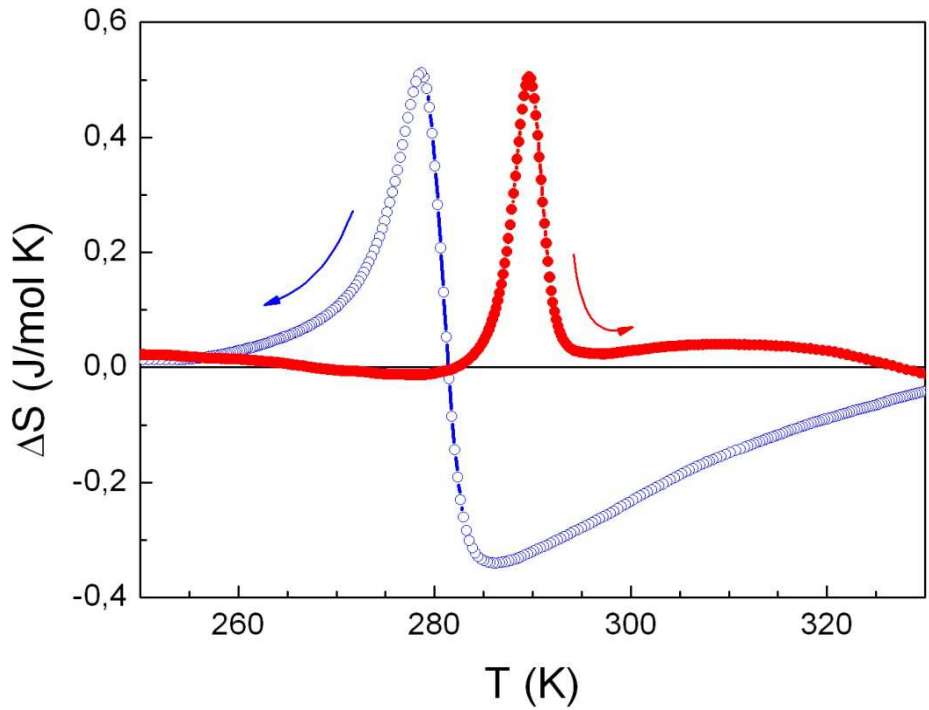

Figure 5. Temperature dependences of MCE in heating and cooling modes at $\Delta H$=11 kOe estimated using the heat capacity data.

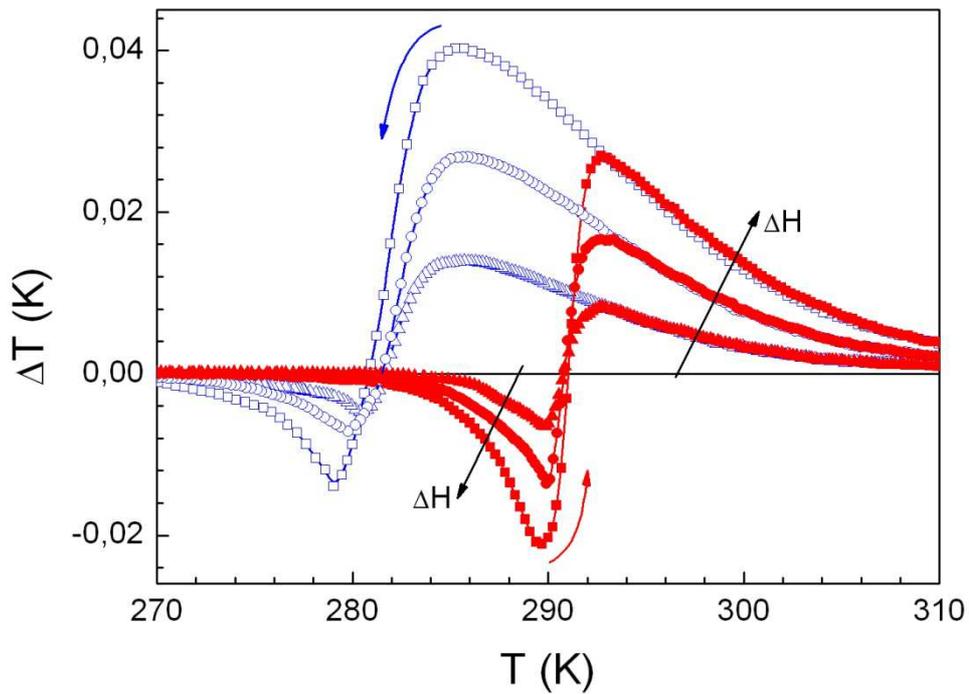

Figure 6. The MCE in $NdBaMn_2O_6$ in heating and cooling modes at magnetic field change 250, 500 and 750 Oe.